\documentclass{article}  
\usepackage{breckenridge}

\usepackage{graphicx}
\usepackage{float}
\usepackage{amssymb}

\frompage{000} \topage{000}                                              %|
\def\rightmark{High $p_T$ identified particles in PHENIX: data {\it vs} theory}
\def\leftmark{David d'Enterria (PHENIX Collaboration)}
 
\title{High $p_T$ identified particles in PHENIX: data {\it vs} theory} 
\authors{
{David d'Enterria$^1$ for the PHENIX Collaboration
}\\[2.812mm]
{\normalsize
\hspace*{-8pt}$^1$ Nevis Laboratories, Columbia University\\ 
Irvington, NY 10533, and New York, NY 10027, USA\\[0.2ex] 
}}
 
\abstract{Two of the most interesting experimental results of 
heavy-ion reactions at RHIC collider energies are in the hard scattering
sector where central Au+Au data show a very different behaviour compared 
to $p+p$ and peripheral Au+Au collisions. The so-called ``high $p_T$ $\pi^0$ 
suppression'' and the ``anomalous'' baryon/meson ratio observed by PHENIX in 
central Au+Au collisions at $\sqrt{s_{NN}}$ = 200 GeV
are reviewed and compared to various theoretical calculations based 
on different strongly interacting medium scenarios. 
}

\keyword{Hard scattering, relativistic heavy-ions, high $p_T$, PHENIX, RHIC} 
\PACS{25.75w}
 
\begin{document}
 
\maketitle
\setcounter{page}{1}

\section{Introduction}

Heavy-ion collisions at RHIC collider energies aim at the study
of QCD matter at extreme energy densities where lattice calculations \cite{latt} predict a transition from hadronic matter 
to a deconfined, chirally symmetric plasma of quarks and gluons (QGP).
The most interesting phenomena observed so far in Au+Au reactions at
$\sqrt{s_{_{NN}}}$ = 130 and 200 GeV, are in the high transverse momentum ($p_T \gtrsim$ 2 GeV/$c$)
sector where the production of hadrons in central collisions shows
subtantial differences compared to more elementary ($p+p$, $e^+e^-$)
systems. One first observation \cite{ppg003,ppg014} is that the high $p_T$ yield of $\pi^0$ %with $p_T\gtrsim$ 4 GeV/$c$ 
in central Au+Au is suppressed by as much as a factor 4--5 compared to $p+p$
and peripheral Au+Au scaled by the number of nucleon-nucleon ($NN$) collisions,
$N_{coll}$. A second observation is that at intermediate $p_T$'s (1 $< p_T <$ 4.5 GeV/$c$) 
no suppression is seen for protons and anti-protons, yielding an ``anomalous'' 
baryon over meson $p/\pi\sim$ 1 ratio \cite{ppg008,ppg015}
much larger than the ratio $p/\pi\sim$ 0.1 -- 0.3 observed in $p+p$ \cite{ISR_ppbar,ISR_pi0} and in $e^+e^-$
jet fragmentation \cite{DELPHI}. Both results point to strong medium effects at work in central 
Au+Au collisions, and have triggered extensive theoretical discussions
based on perturbative \cite{vitev,ina,mueller,arleo,carlos,levai,xnwang} or ``classical'' \cite{dima} QCD. 
Most of the studies on the high $p_T$ suppression are based on the prediction \cite{Gyu90} 
that a deconfined and dense medium would induce multiple gluon radiation off 
the scattered partons, effectively leading to a depletion of the high-$p_T$ hadronic 
fragmentation products (``jet quenching''). Alternative interpretations %for the observed deficit 
have been also put forward based on initial-state gluon saturation 
effects \cite{dima}, or final-state hadronic reinteractions \cite{gallmeister}. 
The different behaviour of baryons and mesons at moderately high $p_T$'s 
has been interpreted, among others, in terms of ``parton recombination'' 
effects \cite{recomb}. It is the purpose of this paper to present the $\sqrt{s_{NN}}$, 
$p_T$, centrality, and baryon-meson dependence of the high $p_T$ production measured 
by PHENIX in Au+Au, and compare it to the different proposed physical 
explanations in order to shed some light on the properties of the underlying QCD %strongly interacting 
medium of heavy-ion collisions at RHIC.\\
%(i) final-state parton energy loss (``jet quenching''), 
%(ii) initial-state parton saturation (``color glass condensate''), 
%(iii) final-state parton recombination, and 
%(iv) final-state hadronic energy loss.

PHENIX \cite{nim} is specifically designed to measure penetrating probes 
by combining good mass and particle identification (PID) resolution in a broad momentum range, 
high rate capability, and small granularity. The results presented here concentrate on the
Run-2 measurement of (i) neutral pions \cite{ppg014} reconstructed, via their $\gamma\gamma$
decay, in the electromagnetic calorimeter EMCal (covering $\Delta\eta = 0.7$, 
$\Delta \phi = \pi$), together with (ii) protons and anti-protons \cite{ppg015}
identified using the central tracking system (a multi-layer drift chamber followed by a multi-wire
proportional chamber with pixel-pad readout) plus the time-of-flight detector 
($\Delta\eta = 0.7$ and $\Delta \phi = \pi/8$).
The $\pi^0$ ($p,\bar{p}$) analysis uses $\sim 20\,(30) \times 10^6$ minimum bias events, 
triggered by a coincidence of the Beam-Beam Counters (BBC) and Zero-Degree Calorimeters (ZDC),
with vertex position $|z|<$ 30 cm. % from the nominal interaction point. 
%In order to study the particle production yields in Au+Au as a function of the reaction centrality, and 
%in order to compare the Au+Au and the $p+p$ results, it is important to scale the $p+p$ yields 
The average number of participant nucleons ($N_{part}$) and $NN$ collisions ($N_{coll}$) 
in each Au+Au centrality bin are obtained from a Glauber calculation combined with
the BBC and ZDC responses \cite{ppg014}.

\section{High $p_T$ $\pi^0$ suppression: $\sqrt{s_{NN}}$ and $p_T$ dependence}
  
\begin{figure}[H]%[htb]
%\vspace*{-.3cm}
\includegraphics[height=7.cm]{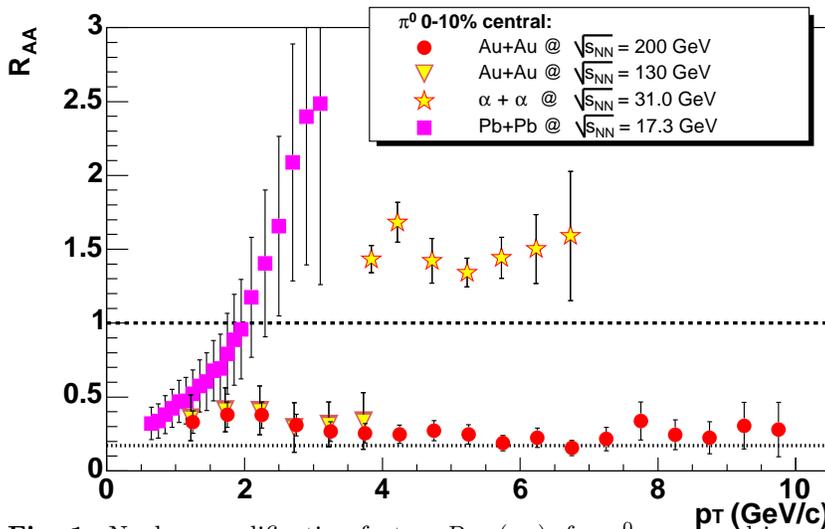}
%\insertplot{RAA_RHIC_ISR_SPS.eps}
\vspace*{-0.5cm}
\caption[]{Nuclear modification factor, $R_{AA}(p_T)$, for $\pi^0$ measured in
central ion-ion reactions at CERN-SPS \cite{wa98_pi0}, CERN-ISR \cite{ISR_pi0}, 
and BNL-RHIC \cite{ppg003,ppg014} energies.}
%: (i) CERN-SPS central Pb+Pb \cite{wa98_pi0} (squares), (ii) CERN-ISR $\alpha+\alpha$ \cite{isr_pi0}
%(stars), (iii) PHENIX-RHIC central Au+Au at 130 GeV (crosses), and (iv) PHENIX-RHIC central 
%Au+Au at 200 GeV (circles).}
\label{fig:R_AA_pi0_syst}
\end{figure}

Details on hadron production mechanisms in $AA$ can be obtained
by investigating their scaling behavior with respect to $p+p$ collisions: 
soft processes ($p_T<$ 1 GeV/$c$) are expected \cite{wnm} to 
scale\footnote{They actually approximately do, see Fig. \ref{fig:R_AA_vs_cent} (right plot, circles) and \cite{shura}.} 
with $N_{part}$, and hard process with $N_{coll}$ \cite{dde_glauber}. 
In order to quantify the medium effects at high $p_T$, one uses the {\it nuclear modification factor} 
given by the ratio of the $AA$ to the $N_{coll}$ scaled $p+p$ invariant yields:
\begin{equation} 
R_{AA}(p_T)\,=\,\frac{d^2N^{\pi^0}_{AA}/dy dp_T}{\langle N_{coll}\rangle\,\times\, d^2N^{\pi^0}_{pp}/dy dp_T}.
\label{eq:R_AA}
\end{equation}
$R_{AA}(p_T)$ measures the deviation of $AA$ from an incoherent superposition 
of $NN$ collisions in terms of suppression ($R_{AA}<$1) or enhancement ($R_{AA}>$1). 
Figure~\ref{fig:R_AA_pi0_syst} shows $R_{AA}$ as a function of $p_T$  for several $\pi^0$ measurements 
in high-energy $AA$ collisions. The PHENIX $R_{AA}$ values for central collisions at 
200 GeV (circles) and 130 GeV (triangles) are noticeably below unity in constrast to the 
enhanced production ($R_{AA}>$1) observed at CERN-ISR \cite{ISR_pi0} (stars) 
and CERN-SPS energies \cite{wa98_pi0} (squares) and interpreted in terms of initial-state 
$p_T$ broadening (``Cronin effect'' \cite{cronin}).

\vspace*{-0.3cm}
The high $p_T$ suppression in central collisions at RHIC is smallest at 2 GeV/$c$ and increases 
to an approximately constant suppression factor of 1/$R_{AA}\approx$~4--5 over $p_T$~=~4 -- 10~GeV$/c$, 
$\sim$30\% above the expectation from $N_{part}$ scaling (dotted line).
The magnitude and $p_T$ dependence\footnote{Corresponding to parton fractional momenta 
$x\approx 2p_T/\sqrt{s}\sim$ 0.02--0.1 at midrapidity.} of $R_{AA}$, is alone inconsistent with 
``conventional'' nuclear effects like leading-twist ``shadowing'' of 
the nuclear PDFs %parton distribution functions
 \cite{eks98,vogt}.\\
%\vspace{-0.1cm}
Different pQCD-based jet quenching calculations % \cite{vitev,ina,levai,xnwang,arleo}, 
based on medium-induced radiative energy loss, can reproduce the {\it magnitude} 
of the $\pi^0$ suppression assuming the formation of a hot and dense partonic system 
characterized by different, but related, properties: 
i) large initial gluon densities $dN^g/dy\sim$ 1000 \cite{vitev}, % 800--1200 \cite{vitev}, 
ii) large ``transport coefficients'' $\hat{q}_0\sim$ 3.5  GeV/fm$^2$ \cite{arleo}, 
iii) high opacities $L/\lambda\sim$ 3--4 \cite{levai}, or iv) effective parton 
energy losses of the order of $dE/dx\sim$ 14 GeV/fm \cite{xnwang}. 
The predicted {\it $p_T$ dependence} of the quenching, however, varies in the different models. 
All models that include the Landau-Pomeranchuck-Migdal (LPM) interference effect \cite{BDMPS,glv} 
predict a slow (logarithmic) increase of $R_{AA}$, not compatible 
with the data over the entire $p_T$ range. Other approaches, such 
as constant energy loss per parton scattering, are also not supported
as discussed in \cite{ina}. Analyses \cite{vitev} which combine LPM jet quenching %together 
with shadowing and initial-state $p_T$ broadening globaly reproduce the 
observed $p_T$ dependence of the $\pi^0$ suppression. However, 
{\it semi-quantitative} estimates of final-state interactions in a dense 
hadronic medium \cite{gallmeister} also seem to yield the same amount of 
quenching as models based on partonic energy loss.

%%%%%%%%%%%%%%%%%%%%%%%%%%%%%%%%%%%%%%%%%%%%%%

\section{High $p_T$ $\pi^0$ suppression at 200 GeV: centrality dependence}

\begin{figure}[H]
%\vspace*{-0.1cm}
\hspace*{-.8cm}
\includegraphics[height=5.9cm]{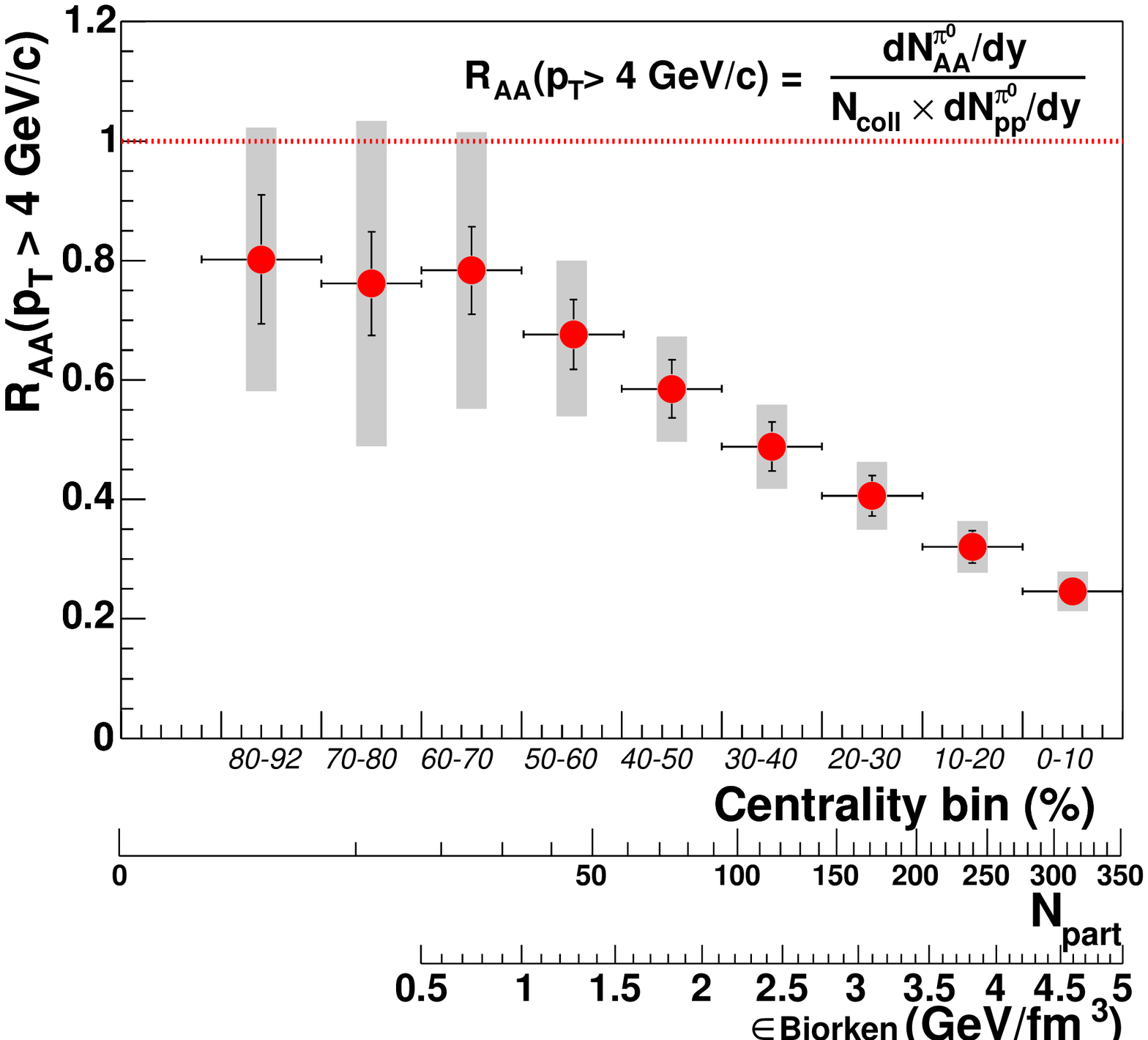}
\includegraphics[height=5.9cm]{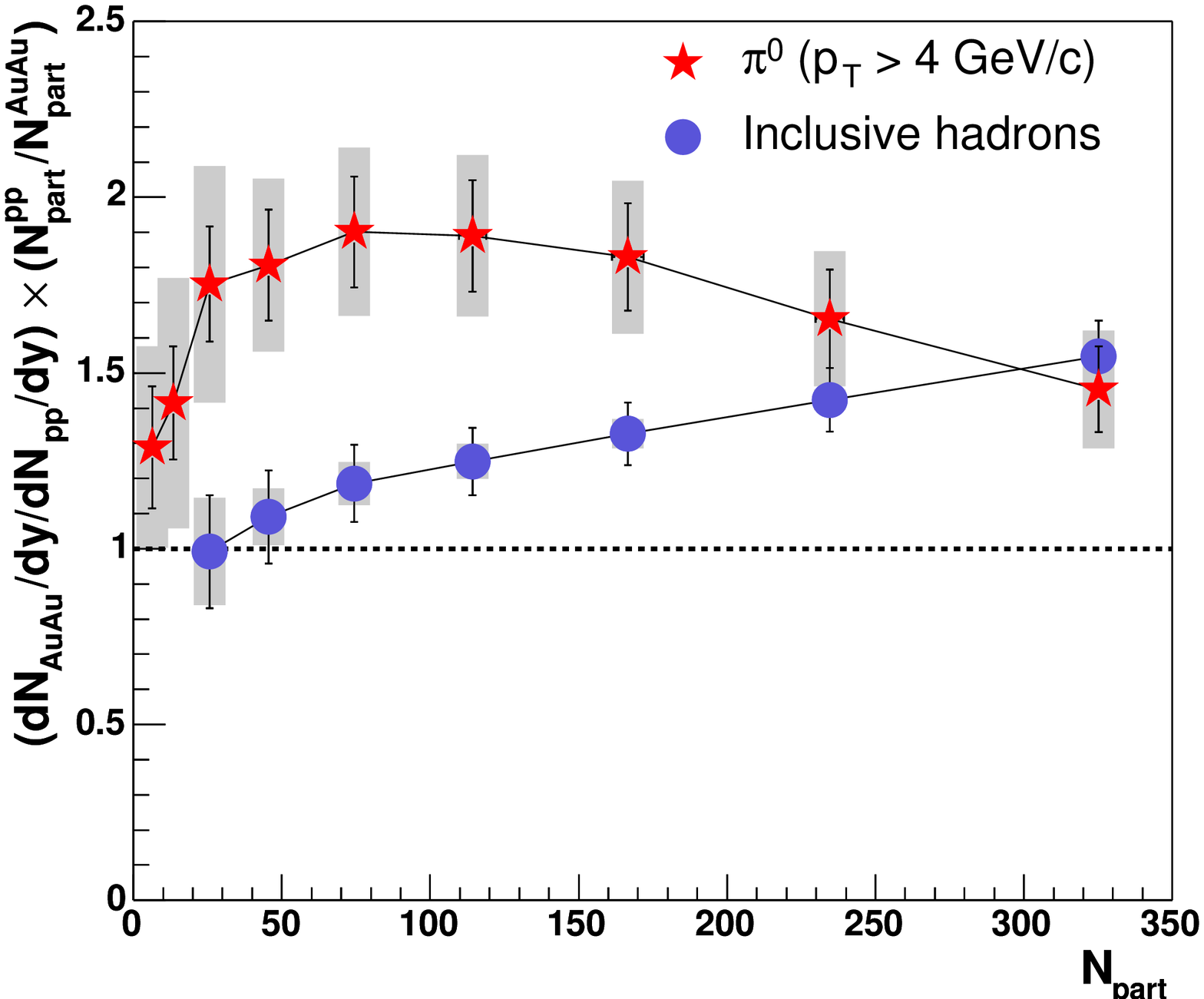}\hfill
%\insertplot{R_AA_pi0_vs_centrality.eps}
%\includegraphics[height=6.0cm]{R_AA_pi0_vs_centrality.eps}\hfill
\hspace*{2cm}
\vspace*{-0.7cm}
\caption[]{{\it Left:} Evolution of the high $p_T$ suppression, $R_{AA}(p_T>$4 GeV/$c$), as a function
of (i) centrality bin, (ii) $N_{part}$, and (iii) ``Bjorken'' energy density. {\it Right:}  In stars 
(circles), the ratio of $Au+Au$ over $p+p$ high $p_T$ $\pi^0$ (inclusive hadron) yields 
normalized by $N_{part}$, as a function of $N_{part}$.}
\label{fig:R_AA_vs_cent}
\end{figure}

In each centrality bin, the value of the high $p_T$ suppression can be
quantified by the ratio of Au+Au over $N_{coll}$-scaled $p+p$ yields integrated, e.g., 
above 4 GeV/{\it c}. %!l: $R_{AA}(p_T>$4 GeV/$c) = (dN_{AuAu}/dy)/(dN_{pp}/dy\times N_{coll})$.
Fig.~\ref{fig:R_AA_vs_cent} (left) shows $R_{AA}(p_T>$4 GeV/$c$) as a function of:
(i) centrality, (ii) $N_{part}$, and (iii) ``Bjorken'' energy density\footnote{
%Obtained using the `Bjorken estimate'': $\epsilon_{Bj} = 1/(\tau_0\,A_{eff}) \times dE_T/dy$ with $\tau_0$ = 1 fm/$c$. 
For each centrality class, $\epsilon_{Bj} = 1/(\tau_0\cdot A_{eff}) \times dE_T/dy$ (``Bjorken estimate'', 
with $\tau_0$ = 1 fm/$c$), where $dE_T/dy$ is the average transverse energy 
measured by PHENIX at $y = 0$ \cite{shura}, and $A_{eff}$ is the overlap area
of the colliding ions given by a Glauber Monte Carlo.} 
$\epsilon_{Bj}$. Within errors, high $p_T$ production in 60--92\% peripheral collisions is 
consistent with $NN$ scaling ($R_{AA}\approx$ 1). A departure from this scaling (at a 2-sigma level) is apparent
for the 50--60\% centrality corresponding to $b\approx$ 11 fm, 
$\sim$ 50 participants, or $\epsilon_{Bj}\approx$ 1.2 GeV/fm$^{3}$.
For more central reactions the suppression increases steadily up to a factor 4--5. 
Whether there is an abrupt or gradual suppression pattern cannot be ascertained 
with the present uncertainties of the data. % in the peripheral range. 
However, {\it assuming} that there is a flat $N_{coll}$ scaling in the most peripheral region,
the departure from $N_{coll}$ scaling for the 50--60\% centrality is (perhaps coincidentally)
in the ball-park of theoretical estimates for: (i) the minimal amount of nucleons  
needed to obtain parton ``percolation'' at RHIC energies ($N_{part}\sim$ 80) \cite{satz}, and
(ii) the expected ``critical'' QCD energy density ($\epsilon_{crit}\approx$ 1 GeV/fm$^3$) \cite{latt}.

\vspace*{-0.2cm}
$N_{part}$ (instead of $N_{coll}$) scaling at high $p_T$ is %(mind that $N_{part}^{pp}$ = 2) 
expected in scenarios dominated either by gluon saturation \cite{dima} or by surface emission 
of the quenched jets \cite{mueller}. Fig.~\ref{fig:R_AA_vs_cent} (right) shows the centrality dependence
of the ratio of $N_{part}$-scaled Au+Au over $p+p$ yields for high $p_T$ $\pi^0$ (stars), 
%The centrality dependence of the high $p_T$ $\pi^0$ multiplicities 
compared to the (clearly different) centrality dependence of the global charged hadron 
multiplicities\footnote{The ratio of inclusive Au+Au 
over p+p multiplicities has been obtained normalizing PHENIX $dN_{AuAu}/dy$ data \cite{shura} by 
the $dN_{p\bar{p}}/dy$ = 2.48 $\pm$ 0.1 measured by UA5 at $\sqrt{s_{_{NN}}}$ = 200 GeV \cite{UA5}.} (circles).
Although there is no true participant scaling for high $p_T$ $\pi^0$
($R_{AA}^{part}>$ 1 for all centralities), their production per participant pair %above 4~GeV/{\it c} 
is, within errors, approximately constant over a wide range of intermediate centralities, 
in qualitative agreement with a gluon saturation model prediction \cite{dima}.

%\newpage
\section{High $p_T$ suppression at 200 GeV: baryons {\it vs} mesons}

\begin{figure}[H]
\vspace*{-0.1cm}
\hspace*{-.8cm}
\includegraphics[height=6.cm]{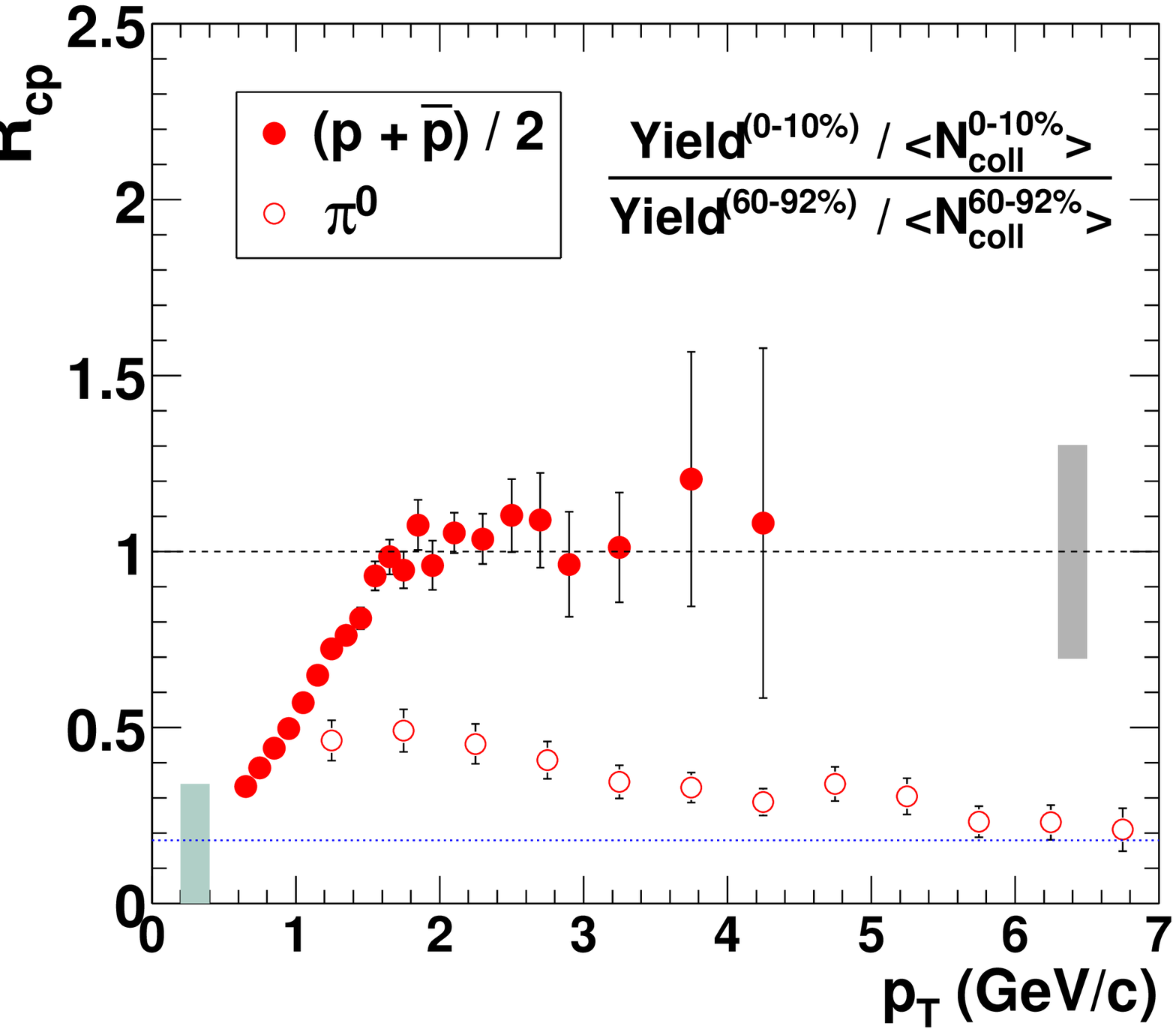}
\hspace*{2mm}\includegraphics[height=5.9cm]{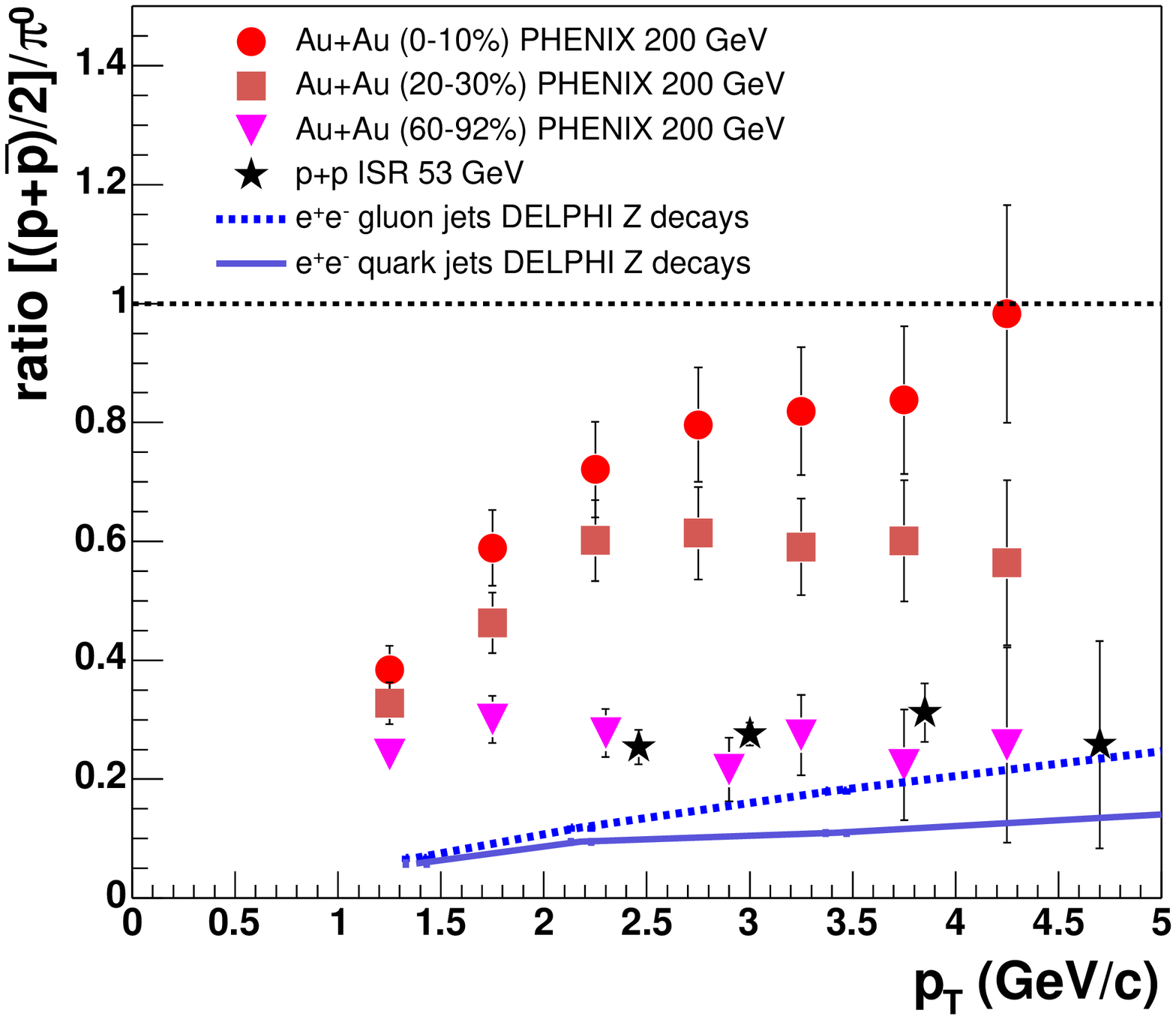}
%\includegraphics[height=5.0cm]{p_pi_ratio.eps}\hfill
%\insertplot{p_pi_ratio.eps}
%\insertplot{Rcp_p_pbar_pi0.eps}
\hspace*{-2cm}
\vspace*{-1cm}
\caption[]{{\it Left}: Ratio of central over peripheral $N_{coll}$ scaled $(p+\bar{p})/2$
(dots) and $\pi^0$ (circles) yields in Au+Au,  as a function of $p_T$. 
{\it Right}: Ratios of  $(p+\bar{p})/2$ over $\pi^0$ versus $p_T$ in: central (dots), 
mid-central (squares), and peripheral (triangles) Au+Au, and in $p+p$ (crosses), and 
$e^{+}e^{-}$ (dashed and solid lines) collisions.}
\label{fig:flavor_dep}
\end{figure}

Figure~\ref{fig:flavor_dep} (left) compares the $N_{coll}$ scaled central to
peripheral yield ratios for $(p+\bar{p})/2$ and $\pi^0$: 
$R_{cp} = (yield^{(0-10\%)} / N_{coll}^{0-10\%}) / (yield^{(60-92\%)} /  N_{coll}^{60-92\%})$. 
Since the %(both the $\pi^0$ and the inclusive charged hadron \cite{}) 
60--92\% peripheral Au+Au spectra scale with $N_{coll}$ when compared 
to the $p+p$ yields \cite{ppg014,ppg015}, $R_{cp}$ carries basically the same information 
as $R_{AA}$. %, Eq. (\ref{eq:R_AA}). 
From 1.5 to 4.5 GeV/$c$ the (anti)protons are not suppressed ($R_{cp}\sim$ 1) 
at variance with the pions which are reduced by a factor of 2--3 in this $p_T$ range.
%Beyond $p_T \simeq$ 1.5 GeV/$c$ all spectra converge to the
%{\emph same} slope and seem to obey $N_{coll}$ scaling in agreement 
%with production due to hard processes in the absence of nuclear effects. 
If both $\pi^0$  and  $p,\bar{p}$ originate from the fragmentation 
of hard-scattered partons that lose energy in the medium, the nuclear 
modification factor $R_{cp}$ should be independent of particle species 
contrary to the experimental result.

Figure~\ref{fig:flavor_dep} (right) shows the ratios of $(p+\bar{p})/2$ over $\pi^0$ as a
function of $p_T$ measured in central (0--10\%, circles), mid-central (20--30\%, squares), 
and peripheral (60--92\%, triangles) Au$+$Au collisions, together with the
corresponding ratios measured in $p+p$ collisions at CERN-ISR energies \cite{ISR_ppbar,ISR_pi0} 
(crosses) and in gluon and quark jets from $e^{+}e^{-}$ collisions \cite{DELPHI}
(dashed and solid lines resp.). Within errors, peripheral Au+Au results are compatible 
with the $p+p$ and $e^{+}e^{-}$ ratios, but central Au+Au collisions have a $p/\pi$ ratio
$\sim$4--5 times larger. Such a result is at odds with standard perturbative production 
mechanisms, %holding in both channels,
%which are well described by a universal fragmentation function
%independent of the colliding system. 
%For a standard perturbative description to hold 
since in this case the particle ratios $\bar{p}/\pi$ and $p/\pi$ 
should be described by a universal fragmentation function independent of the 
colliding system, which favors the production of the lightest particle.
One explanation of this observation is based on
the recombination (or coalescence), rather than fragmentation, of quarks \cite{recomb}.
In this picture the partons from a thermalized system recombine and with the addition of
quark momenta, the soft production of baryons extends to much larger
values of $p_T$ than that for mesons.  In this scenario, the effect is
limited to $p_T < 5$\,GeV, beyond which fragmentation becomes the
dominant production mechanism for all species\footnote{This indeed seems to be confirmed
by the measured $h/\pi^0\sim$~1.6 above $p_T\sim$ 5 GeV/$c$ in central and peripheral Au+Au 
 \cite{ppg015} which is consistent with that measured in $p+p$ collisions and indicates
that for higher $p_T$'s the baryon yields approach the (suppresed) pion scaling}.
Alternative explanations based on medium-induced difference in the formation 
time of baryons and mesons, different ``Cronin enhancement'' for protons
and pions, or ``baryon junctions'' are considered in \cite{ppg015}.

%Beyond $p_T \approx 4.5$~GeV/$c$, the identification of charged particles is not yet
%possible with the current PHENIX configuration, however the measured $h/\pi^0\sim$~1.6
%ratio above $p_T\sim$ 5 GeV/$c$ in central and peripheral Au+Au 
%is consistent with that measured in $p+p$ collisions \cite{ppg015}.
%This indicates that for high $p_T$ the baryon yields should approach the 
%(suppresed) pion scaling, thus limiting the observed baryon enhancement 
%in central Au$+$Au collisions to the intermediate transverse momenta 
%$p_T \lesssim$ 5 GeV/$c$.

\section{Conclusions}

PHENIX has measured $\pi^0$ and $p,\,\bar{p}$ at mid-rapidity up to 
$p_T$ = 10~GeV$/c$ and 4.5~GeV$/c$ respectively, in different 
centrality bins of Au+Au collisions at $\sqrt{s_{_{NN}}}$ = 200~GeV, permiting
a detailed comparative study of the properties of high $p_T$ particle production in 
high-energy heavy-ion collisions as a function of $\sqrt{s}$, $p_T$, %transverse momentum, 
centrality, and particle composition. The spectral shape and  invariant yields 
of {\it all} identified particles in {\it peripheral} reactions are consistent 
with those of $p+p$ reactions scaled by the number of inelastic 
$NN$ collisions, in agreement with pQCD expectations.
For $\pi^0$, a departure from ``$NN$ scaling'' at a 2-sigma level is apparent
for the 50-60\% centrality class corresponding to $\sim$ 50 participant nucleons,
$b\approx$ 11 fm, or $\epsilon_{Bj}\approx$ 1.2 GeV/fm$^{3}$.
For more central reactions the suppression %with respect to the binary scaling expectation 
increases steadily up to a factor 4-5. The existence or not of an abrupt or gradual suppression 
pattern %from the most peripheral down to the most central reactions 
cannot be ascertained within the present uncertainties. The magnitude of the deficit in 
central collisions can be reproduced by pQCD-based parton energy loss calculations 
in an opaque medium, but its $p_T$ and centrality dependence 
puts strong constraints on the details of the energy loss and the properties of the medium.
At variance with pions, %protons and antiprotons are not suppresed in central collisions.
a large $p,\bar{p}$ production, which increases from peripheral to central collisions, 
is observed in the range $ 1.5 < p_T < 4.5 $\,GeV/$c$. In this $p_T$ range, the $p$ and 
$\bar{p}$ yields show ``$NN$ scaling'' independent of centrality, resulting in a much larger
baryon over meson ratio than observed in $p+p$ and $e^+e^-$. % collisions.
%as expected for hard-scattering in the absence of nuclear effects. 
%This baryon enhancement seems to be limited to %transverse momenta 
%$p_T < 4.5$\,GeV/$c$, as deduced from the measurement of the ratio of non-identified 
%charged hadrons to $\pi^0$. 
This observation can be explained in terms of parton energy loss plus recombination effects in a thermalized medium.
Stronger experimental constraints on the properties of the underlying QCD medium produced in 
$AA$ reactions will be available with the data of the recent $d+Au$ run at RHIC, where final-state 
medium effects are absent.

\vfill\eject
\end{document}